\documentclass[aps,twocolumn,showpacs,showkeys]{revtex4}
\usepackage{graphicx}
\usepackage{amssymb}

\begin{document}
\setcounter{page}{1}
\title[]{Passive Sliders and Scaling: from Cusps to Divergences}
\author{Apoorva \surname{Nagar}}
\email{apoorva@kias.re.kr}
\thanks{}
\affiliation{Department of Physics, Korea Institute for Advanced Study, Seoul, Korea}
\author{Mustansir \surname{Barma}}
\affiliation{Department of Theoretical Physics, Tata Institute of Fundamental Research, Mumbai, India}
\date[]{Received January 5 2004}

\begin{abstract}
The steady state reached by a system of particles sliding down a fluctuating surface has interesting properties. Particle clusters form and break rapidly, leading to a broad distribution of sizes and large fluctuations. The density-density correlation function is a singular scaling function of the separation and system size. A simple mapping is shown to take a configuration of sliding hard-core particles with mutual exclusion (a system which shows a cusp singularity) to a configuration with multiparticle occupancy. For the mapped system, a calculation of the correlation function shows that it is of the same scaling form again, but with a stronger singularity (a divergence) of the sort observed earlier for noninteracting passive particles.
\end{abstract}


\maketitle

\section{INTRODUCTION}
Driven diffusive systems consist of many particles, with individual
particles undergoing a diffusive motion apart from being influenced by
interparticle interactions and external forcing, which drives the
system into a nonequilibrium state. Examples range from stirred fluids
on the one hand, to current-carrying systems, such as vehicular and
pedestrian traffic, on the other.  Attempts at theoretical modeling of
such systems range from setting up and trying to solve continuum
equations like the Navier-Stokes equations for fluids to studying
lattice models like the asymmetric exclusion process, a simple
model for directed motion of particles with an exclusion constraint.
The theoretical challenge is to describe the macroscopic properties of
these nonequilibrium systems in the absence of a general prescription
that specifies the weights of microscopic configurations in the
steady state, akin to the Boltzmann-Gibbs prescription for equilibrium
statistical mechanics.

The coupling of two or more driven diffusive systems to each other can
give rise to complex and interesting behavior.  This is so even when
the coupling is unidirectional; {\it i.e.}, one of the driven fields
evolves autonomously and drives the other (passive) field.  An example
is the problem of passive scalars, like ink or dye, advected by the
streamlines of a stirred fluid \cite{falkovich1}. The nontrivial
nature of the passive scalar problem arises from the fact that besides
being driven by the fluid, the passive particles also diffuse; this
allows passive particles to jump from one advecting streamline to
another, leading to an intricate behavior of the passive density. Of
course, the nature of the driving field is of great importance for the
ultimate distribution of the scalar.  For instance, passive particles
driven by incompressible fluids (e.g., ink in water) tend to spread out
and mix in the large-time limit.  However, if the fluid in
question is compressible, the behavior can change drastically, and it
is possible for the particles to cluster together in dynamic clumps
rather than reach a homogeneous state \cite{gawedzki}. The
instability of a homogeneous state to clustering or clumping has been
discussed earlier, both in the context of driving by compressible
fluids and separately by allowing for the inertia of driven particles,
which allows them to deviate from strictly following the streamlines
\cite{gawedzki,maxey,balkovsky}.  It is clearly of interest to
characterize the steady states in such situations, where there is a broad
distribution of cluster sizes under conditions of rapid making and breaking of individual clusters.

This sort of dynamical steady state, in which particle clusters
constantly form and break, has recently been studied in another
related context, namely, particles sliding down fluctuating surfaces,
which themselves are driven systems
\cite{das1,das2,manoj,nagar1,nagar2,chatterjee,barmasingular}.  In a
qualitative sense, the state is quite different for the cases of
noninteracting passive particles \cite{nagar1,nagar2} and passive
particles with mutual hard-core exclusion interactions
\cite{das1,das2,manoj,chatterjee}.  In the former case, there is a
large degree of clumping accompanied by strong fluctuations, as large,
concentrated clusters can form and break; this state is referred to as
a strong clustering state (SCS) \cite{nagar2}.  In the latter case,
mutual exclusion prevents pile-ups of particles at the same spatial
location.  The state turns out to have long-range order as in phase-ordered systems familiar from equilibrium contexts. However, unlike
equilibrium systems, fluctuations in this case remain very strong even
in the thermodynamic limit hence the appellation
fluctuation-dominated phase ordering (FDPO)\cite{das2}.  At a
quantitative level, the differences between the two cases are captured
by the two-point density-density correlation function.  Numerical
simulations show that for SCS, as well as for FDPO, the correlation
function is a scaling function of separation and system size.
However, the scaling functions are quite different in the two cases,
being characterized by different sorts of singularities for small
values of the scaling argument: a divergence for the case of SCS and
a cusp singularity for the case of FDPO.

Beyond the numerical results, it is useful to have analytical treatments for simplified models, in order to explicitly demonstrate the existence of scaling and singularities of the scaling function.  Such a treatment was carried out for FDPO by considering the properties of a coarse-grained depth model of the surface \cite{das1,das2,manoj,chatterjee}.  The resulting scaling function shows a cusp singularity. The principal new result reported in this paper is that a simple mapping takes a configuration of the FDPO steady state in such a model to a configuration that is of the SCS variety.  This allows an explicit calculation of correlation functions and a demonstration of scaling with a divergent scaling function.

The paper is organized as follows: In Section II, we discuss lattice
models of driven, passive sliders for both noninteracting and
interacting cases and review the scaling properties for the two-point
correlation function, {\it vis-\`a-vis} the singular behavior characterizing
SCS and FDPO.  In Section III, we construct a variant of the
coarse-grained depth model and demonstrate that the two-point
correlation function exhibits a cusp.  We then consider the effect of
a mapping from configurations of this model (with at most one particle
per site) to a model with multiple occupancies, and demonstrate a
divergence of the scaling function in the new model.  Thus, the cusp
singularity --- the hallmark of FDPO --- contains the seeds of a
divergence in the mapped model, the characteristic of SCS.

\section{SLIDING PARTICLES ON FLUCTUATING SURFACES: A SURVEY}
In this section, we summarize recent work on the problem of
passive particles sliding under gravity on stochastically evolving
surfaces
\cite{das1,das2,manoj,nagar1,nagar2,chatterjee,barmasingular,drossel,drossel1,bohr,chin,manoj2}.
The surfaces under consideration are
taken to evolve
according to the Kardar-Parisi-Zhang (KPZ) and the Edwards-Wilkinson
(EW) dynamics. Apart from the effect of gravity, the particles also
have a random noise acting on them. The nature of the interaction between
particles is an important consideration and has a significant impact
on the behavior of
the system. Two cases were considered --- hard core
repulsion and no interaction at all, {\it i.e.}, noninteracting particles. In both cases, one sees a clustering of particles and finds
strong fluctuations. However, the nature of clustering depends strongly
on whether we have hard-core repulsion, which allows a
finite occupancy, or no interaction, which allows for arbitrarily high
particle occupancies.

\subsection{Noninteracting Particles}

Let us first consider the KPZ equation for an evolving surface:

\begin{eqnarray} {\partial h \over \partial t} = \nu \nabla^{2} h + \frac{\lambda}{2}(\nabla h)^{2}+\zeta_h(\vec{x},t),
\label{kpz}
\end{eqnarray}
which describes an evolving height field $h(\vec{x},t)$. $\zeta_h$ is
a Gaussian white noise satisfying $\langle \zeta_h (\vec{x},t)
\zeta_h(\vec{x}',t')\rangle = 2D_h \delta^d(\vec{x} -
\vec{x}')\delta(t - t')$. This equation contains the nonlinear term $
\frac{\lambda}{2}(\nabla h)^{2}$, which breaks the $h \rightarrow -h$
symmetry and allows for the possibility of the surface moving in the
direction of particle motion or against it.  The transformation
$\vec{v}=-\nabla h$ maps the above equation (with $\lambda=1$) to the
Burgers equation for a compressible fluid, $\vec{v}$ being the
velocity field of the fluid. The problem of sliding particles on
surfaces then becomes the passive scalar problem of fluid dynamics,
which describes the motion of an advected field in a stirred
fluid. \\

Consider noninteracting particles that slide on the fluctuating
surface described by Eq. 1. These particles sense the local
slope and tend to move downwards, as if subject to gravity. In
addition to this downward movement, the particles are also subject to
random white noise. This problem was first studied by Drossel and
Kardar~\cite{drossel,drossel1}. A useful approach to study this
coupled surface-particle system is to study
a lattice model by using Monte-Carlo
simulations~\cite{nagar1,nagar2,drossel,drossel1}. The model of Refs.~\cite{nagar1} and~\cite{nagar2} consists
of a flexible one-dimensional lattice in which particles reside on
sites while the links or bonds between successive lattice sites are
also dynamical variables that denote local slopes of the surface. The
asymmetry of the KPZ dynamics allows for two kinds of dynamics, namely,
advection and anti-advection, with particles moving in the
direction and against the direction of surface motion,
respectively. The possibility of different time scales of particle and
surface motion was modeled by using the
ratio $\omega$ of the particle to surface update rates. In particular, the
limit $\omega \rightarrow 0$, with $L$ held fixed, corresponds to the
adiabatic limit of the problem where particles move on a static, disordered
surface, and the steady state is a thermal equilibrium state. Exact
analytic results can be obtained in this limit \cite{nagar2}.\\

We will begin by describing the results for the $1-d$ KPZ advection case
described above. While various aspects of the steady state have been
studied~\cite{drossel,drossel1,nagar1,nagar2,bohr,chin,manoj2}, we
restrict our discussion here to the relevant static
quantities. For finite values of $\omega$, Monte-Carlo simulations
were used to evaluate the two point density-density correlation
function $G(r,L) \equiv \langle n_i n_{i+r}\rangle_L$ where $n_i$ is
the number of particles at site $i$. Numerical data for
various system sizes $L$ were shown to be consistent with the scaling
form
\begin{eqnarray}
G(r,L) \sim \frac{1}{L^{\mu}} Y\left({\frac{r}{L}}\right).
\label{correlation}
\end{eqnarray}
Here, $\mu \simeq {1/2}$, and the scaling function $Y(y)$ has a power
law divergence $Y(y) \sim y^{-\nu}$ as $y \rightarrow 0$, with $\nu
\simeq 3/2$.\\

The divergence of the scaling function indicates a strong clustering
of particles while the scaling with system size implies that there are
particle clusters separated from each other on the scale of
the system size. This scaling and divergence are the defining features
of a new kind of steady state --- the strong clustering state or
SCS. Further, the system shows strong fluctuations in the steady state.
These were characterized using the variance $\Sigma^2$ of the fraction
of sites ${\cal{N}}_n/L$ with occupancy $n$. We found that in the
limit $L \rightarrow \infty$, the ratio $\Sigma/{\langle {\cal{N}}_n/L
\rangle}$ approaches a constant. This is to be contrasted with a
normal, self-averaging system where this ratio vanishes in the limit
$L\to \infty$.\\

Let us now turn to the limiting adiabatic case, $\omega \rightarrow 0$,
corresponding to an equilibrium system of particles at inverse
temperature $\beta$ distributed on a disordered, stationary
surface.  Relevant quantities were calculated by
averaging over all surface configurations, as in the Sinai
model~\cite{sinai}. For the KPZ equation in one dimension, the
distribution of heights in the stationary state is described by ${\rm
Prob} [\{h(r)\}]\propto \exp\left[-{1\over {2}}\int
h^2(r')dr'\right]$. Thus, any stationary configuration can be thought
of as the trace of a random walker in space evolving via the equation
$dh(r)/dr = \xi(r)$, where the white noise $\xi(r)$ has zero mean and
is delta correlated, $\langle \xi(r)\xi(r')\rangle = \delta
(r-r')$. This is exactly the surface considered in the Sinai
model. The probability $\rho (r) \equiv n_{r}/L$ of finding the
particle at position $r$ is given by $\rho(r)= \exp[-\beta h(r)]/Z$
with the partition function $Z=\int_0^L \exp[-\beta h(r')]dr'$. One
can then calculate the correlation function $ G(r,L)/L^2 = \langle
\rho(r_0)\rho(r+r_0) \rangle$ by following the calculation of Comtet and
Texier~\cite{comtet}. In the scaling limit, $r\to \infty$, $L\to
\infty$ with the ratio $y=r/L$ fixed, one finds $G(r,L)\sim L^{-1/2}
Y(r/L)$, where the scaling function $Y(y)$ diverges near the origin as
a power law with a power $3/2$. Surprisingly, this equilibrium
result reproduces very well the scaling exponents and scaling
functions found for the correlation function in the strongly
nonequilibrium case $\omega = 1$. \\

The phenomenon of clustering and SCS is not restricted to the KPZ
advection case. One can also consider other driving surfaces --- the
Edwards-Wilkinson (EW) surface where the nonlinear term of
Eq.~(\ref{kpz}) is absent or the KPZ anti-advection
case where the particles move opposite to the KPZ surface
motion. In both of these cases, the steady state was seen to be an
SCS with the same scaling form as in
Eq.~(\ref{correlation}), but with different
exponents~\cite{nagar1,nagar2}. We found that $\mu = 0$ in both
these cases while $\nu \simeq {1/3}$ and $2/3$ for the
KPZ anti-advection and the EW cases, respectively. These values
indicate clustering is less pronounced than in the KPZ advection
case.\\

To summarize, the system of noninteracting particles sliding on
fluctuating surfaces shows interesting behavior with a high degree
of clustering of particles and very large fluctuations in the
distribution of particles from one configuration to another.
The results agree very well with results
for an equilibrium model with quenched disorder, suggesting that the
action of nonequilibrium surface fluctuations is similar to that
of temperature in the equilibrium problem.

\subsection{Particles Interacting by Hard-core Repulsion}

We now consider particles that are again driven by fluctuating
surfaces as in the previous section, but which have a hard-core
interaction amongst themselves. This problem has been well studied, and
many aspects are
understood~\cite{das1,das2,manoj,chatterjee}. We will concern
ourselves here again with the static properties. As for
noninteracting particles, Monte-Carlo simulations were performed to
study steady-state characteristics ~\cite{das1,das2}. The
dynamical rules for the Monte-Carlo were similar to those discussed
above, but with the
additional restriction that a particle could not move to an already
occupied site. The occupancy is described by an Ising variable
$\sigma_i$ with value $-1$ when a given site
$i$ is unoccupied and $+1$ when it is occupied. The
number of particles is taken to be $L/2$.\\

The quantity of interest is the two-point correlation function
 $C(r,L) \equiv \langle \sigma_{i} \sigma_{i+r}
\rangle$. Numerical simulations show that $C$ is a scaling function of $r$
And $L$:

\begin{eqnarray}
C(r,L) \approx m^2\left[1-a\left(\frac{r}{L}\right)^{\alpha}\right]
\label{fdpocorr}
\end{eqnarray}
as $r/L \rightarrow 0$. The scaling function shows a cuspy fall from a
finite intercept, with cusp exponents $\alpha \simeq 0.25$ for
driving by a KPZ surface and $\alpha \simeq 0.5$ for an EW
surface. The value of the intercept is a measure of long-range order
\cite{das2,barmasingular}, and the system can be thought of as a phase-ordered system similar to a conserved-spin Ising system. The
difference is that in this case, the scaling function shows a cusp
rather than the linear Porod law decay ($\alpha=1$) characteristic of
regular phase-ordered systems, implying that there are no sharp
interfaces between phases. The other feature of the system is the
occurrence of strong macroscopic fluctuations, characterized by using the
lowest wave-vector Fourier components of the density profile
\cite{das2,barmasingular}, thus the name fluctuation dominated phase
ordering (FDPO) for this sort of state.  The clustering of particles
in FDPO is milder than that for SCS, an outcome of interparticle
interactions. In $2-d$, the steady state of particles sliding on a KPZ
surface was found to be of the FDPO variety, too \cite{manoj}. \\

To characterize FDPO analytically, simpler models known as
Coarse-grained Depth (CD) models were
defined in Refs.~\cite{das1} and~\cite{das2}. In the CD models, one considers an evolving
surface, and for each surface configuration, one places an
imaginary cut or reference line, below which all sites are
occupied ($\sigma_{i}=1$) and above which
all sites are empty ($\sigma_i=-1$). One can then
compute the correlation function as before.  Different
prescriptions for choosing the reference level, discussed below, define
various kinds of CD models (CD1, CD2, CD3, ...) ~\cite{das2}. The CD1
model turns out to have an uninteresting steady state \cite{das2} and will not be discussed here.  The CD2 and CD3 models are discussed in the next
paragraph while the CD4 model is defined in Section III.
Analytical results can be obtained for CD models and allow
demonstration of FDPO behavior, with correlation functions showing
scaling, as for the sliding particles discussed above. The CD models
can be thought of
as the very-low-temperature limit of the sliding particle model, where the
particles find the deepest empty sites and occupy them up to a prescribed height, in the adiabatic limit of a frozen surface configuration. As in the case of SCS, this equilibrium, disordered system describes the nonequilibrium
FDPO state very well.\\

For the CD2 model, one considers the cut to be always at the
height of the site $i=0$. As the
configurations of a $1-d$ KPZ or an EW surface can be thought of as the
trajectories of a random walk, the
length of successive stretches of sites above the cut ($\sigma=1$) and
below the cut ($\sigma=-1$) are distributed in the same way as the
first returns to the origin of a random walk. Thus, the probability
distribution $P(l)$ for the length $l$ of the stretches of occupied
and unoccupied sites is given by $P(l) \sim l^{-3/2}$. One can, thus,
calculate the correlation function $C(r)$ by using the fact that
successive intervals of occupied sites (up spins) and unoccupied sites
(down spins) are distributed independently of each other and according
to a power law. We found that $C(r)$ had the same form as in
Eq.~(\ref{fdpocorr}) above with $\alpha=1/2$, which matched very well
the numerical result for the EW surface with sliding
particles. The other model considered was the CD3 model where the
reference line is taken at the level of the instantaneous average
height. The distribution of the lengths of spin up/down clusters was
computed using Monte-Carlo simulations, $P(l) \sim l^{-\theta}$ with
$\theta \simeq 3/2$, as for the CD2 model. To calculate the
correlation function analytically, one can make the approximation that
successive clusters are distributed independently of each other ---
the independent interval approximation (IIA)~\cite{majumdar}. Using
the IIA, the correlation function was found once again to behave as in
Eq.~(\ref{fdpocorr}) with $\alpha=2-\theta=1/2$, as in the CD2
model. This result was verified by numerical simulations.\\

To summarize, the FDPO steady state for hard-core interacting
particles sliding on a fluctuating surface shows
clustering of particles and strong fluctuations. The clustering was
characterized by a cusp in the scaling function describing
the correlation function. One can understand these results for the
nonequilibrium model by studying the simpler CD
models, which correspond to filling a disordered landscape up to a
prescribed level.

\section{Mapping from single-particle to multiparticle occupancies:
from FDPO to SCS}

As we have seen, the simple CD models gave considerable insight into the
nonequilibrium FDPO state.  An analytic treatment was possible as the
cluster size distribution could be connected to the two-point
correlation function, within the independent interval approximation.  The
result --- a scaling function with a cusp singularity --- is the
hallmark of the FDPO steady state. This leads us to ask:
Is it possible for us to similarly find a simple system that
helps to shed light on the SCS steady state whose characteristic is a divergence
of the scaling function for small argument? \\

We take a clue from a simple mapping that connects the simple
exclusion process to the zero-range process
(ZRP) \cite{ZRP}. The
connection takes a system of particles interacting by hard-core repulsion,
with a maximum occupancy of one particle per site, to a system with no limit on occupancy. We implement a similar mapping on the CD
model and show that the resulting model with multiparticle occupancy
has a divergent scaling function of the SCS variety. \\

The mapping works as follows: for a given CD configuration, every
unoccupied site preceding a cluster of particles is assigned a number of
particles equal to that present in the particle cluster; the particle
cluster itself is erased (Figs. 1(c) and 1(d)). This procedure can be interpreted as shifting the particle clusters from a horizontal to a vertical position and placing this vertical cluster on the previous site. The number of lattice sites in the new model is then equal to the number of empty sites in the CD configuration.  We calculate the two-point density-density correlation function for this
new mapped model and demonstrate $r/L$ scaling with a divergence for small argument --- defining features of the SCS.\\

Let us consider a typical configuration of the CD model with
alternating clusters of particles and holes
(empty sites) as in Fig. 1(c). The length $l$ of these stretches is
distributed as a power law $P(l)\sim l^{-\theta}$, where, for the CD
models under consideration, $\theta = 3/2$. We take the average
particle density to be the same as the average hole density. As
illustrated in Figs. 1(c) and 1(d), each configuration of a CD
model can be mapped to a configuration of a
vertical-CD (henceforth, VCD) model with no limit on the occupancy. The
two-point density-density correlation function in the VCD model is
given by

\begin{figure}[t!]
\includegraphics[width=8.0cm]{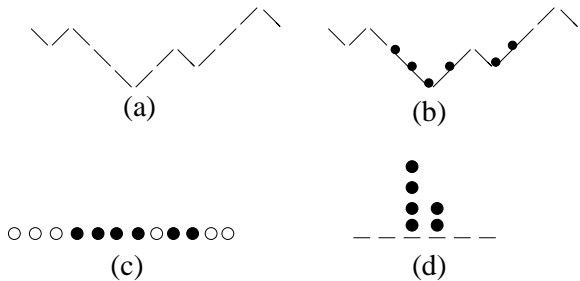}
\caption{Configurations of the CD4 and corresponding VCD model: (a) a typical configuration of a $1-d$ KPZ or EW surface, (b)  $L/2$ particles placed at the deepest sites of the lattice, (c) the resulting CD4 configuration, and (d) the VCD configuration obtained through the mapping.}
\label{mapping}
\end{figure}

\begin{eqnarray}
& &G(r,L) \equiv \langle n(i)n(i+r) \rangle_L\\ \nonumber
&=& \int_{0}^{L}\int_{0}^{L} n(i)n(i+r)P^{\ast}(n(i),n(i+r)) dn(i)dn(i+r).
\end{eqnarray}
Here, $n(i)$ and $n(i+r)$ are the numbers of particles at sites $i$ and
$i+r$. The angular brackets denote an
average over configurations, and $P^{\ast}(n(i),n(i+r))$ is the joint
probability that there are $n(i)$ particles at site $i$ and $n(i+r)$
particles at site $i+r$. Now, $P^\ast$ is given by

\begin{eqnarray}
P^{\ast}(n(i),n(i+r))=P_1(i)Q(n(i))P_2(r)Q(n(i+r)),
\end{eqnarray}
where $P_1(i)$ is the probability that a given site $i$ is occupied
and $P_2(r)$ is the probability that the site at a distance $r$ from
site $i$ is occupied, given that site $i$ is occupied. $Q(n(j))$ is
the probability that the occupancy of site $j$ is $n(j)$, given
that it is occupied.\\

The probability $P_1(i)$ can easily be calculated from the lattice model by evaluating the average number of occupied sites divided by the system size. The average number of occupied sites can be calculated by dividing the system size by the average length $\langle l \rangle$ of the particle clusters in the original CD model,

\begin{eqnarray}
P_1(i)=\frac{L}{\langle l \rangle}/L=a_1L^{\theta-2}
\end{eqnarray}
because $\langle l \rangle$ can be shown to be proportional to $L^{2-\theta}$ by using $\langle l \rangle = \int l P(l) dl$ in the limit of large $l$. Here, $P(l) \sim l^{-\theta}\Theta(L-l)$ is the probability
distribution for the length of the clusters in the CD model and $1 <
\theta < 2$. The $\Theta$ function enforces a cutoff at the system size.\\

We now calculate $P_2(r)$, the probability that site $i+r$ is occupied
given that site $i$ is occupied. Consider the segment of length $r$
following site $i$ in the VCD model. This segment is composed of
$n$ consecutive hole segments in the underlying CD model, where $n$ is
a number between $1$ and $r$. Thus, $P_2(r)$ is the same as the
probability that the length of these $n$ segments in the underlying CD
model adds up to exactly $r$,

\begin{eqnarray}
P_2(r)=\sum_{n=1}^{r}p_n^{*}(r),
\end{eqnarray}
where

\begin{eqnarray}
p_n^{*}(r)&=&\int_0^rdl_1\int_{l_1}^{r}dl_2\int_{l_2}^{r}dl_3....
\int_{l_{n-2}}^{r}dl_{n-1} \times\\\nonumber
& &P(l_1)P(l_2-l_1)P(l_3-l_2)....P(r-l_{n-1}).
\end{eqnarray}
Note that we have gone to a continuum description because we are working
with separations much larger than the lattice constant. Proceeding
as in Ref.~\cite{das2}, we define the Laplace transform of a function $f(x)$
as $\tilde{f}(s)=\int_0^{\infty}dx e^{-sx} f(x)$ and take the Laplace
transform on both sides of Eq. (8), yielding

\begin{eqnarray}
\tilde{p}_n^{*}(s)=\tilde{P}(s)^n,
\end{eqnarray}
where $\tilde{p}_n^{*}(s)$ and $\tilde{P}(s)$ are the Laplace
transforms of $p_n^{*}(r)$ and $P(r)$, respectively. Thus,

\begin{eqnarray}
\tilde{P}_2(s)=\sum_{n=1}^{r}p_n^{*}(r)=\frac{\tilde{P}(s)-\tilde{P}(s)^{r+1}}{1-\tilde{P}(s)}
\end{eqnarray}
where $\tilde{P}_2(s)$ is the Laplace transform of $P_2(r)$. In
the limit of large $r$, we have

\begin{eqnarray}
\tilde{P}_2(s)=\frac{\tilde{P}(s)}{1-\tilde{P}(s)},
\end{eqnarray}
and in the range $1/L \ll s \ll 1$, we can expand $\tilde{P}(s)\approx
1-bs^{\theta-1}$, which gives

\begin{eqnarray}
\tilde{P}_2(s)=\frac{1-bs^{\theta-1}}{bs^{\theta-1}} \approx
\frac{1}{bs^{\theta-1}}.
\end{eqnarray}

Taking the inverse Laplace transform gives

\begin{eqnarray}
P_2(r)=a_2r^{\theta-2}.
\end{eqnarray}
Since our mapping simply flips the particles from a horizontal to a
vertical position, $Q(n(i))\sim n(i)^{-\theta}\Theta(L-n(i))$ and
$Q(n(i+r)) \sim n(i+r)^{-\theta}\Theta(L-n(i+r))$. Thus, finally,

\begin{eqnarray}
G(r,L)&=&ar^{\theta-2} L^{\theta-2}
\int_{\epsilon}^{L-\epsilon}\int_{\epsilon}^{L-\epsilon}
x^{-\gamma}y^{-\gamma} \nonumber \\
&& \Theta(L-x-y)dxdy,
\end{eqnarray}
where $\gamma=\theta-1$, $x \equiv n(i),~y \equiv n(i+r)$, and
$\epsilon$ is a cutoff coming from the finite lattice spacing. Solving
the above integral leads to an expression involving the Gauss
Hypergeometric function $_2F_1(a,b;c;z)$. For large $L$, the leading-order contribution from this integral can be shown to be $L^{4-2\theta}$, implying

\begin{eqnarray}
G(r,L)=a'r^{\theta-2} L^{\theta-2} L^{4-2\theta}=a'(\frac{r}{L})^{\theta-2}.
\label{eqcdsps}
\end{eqnarray}
We, thus, see that the correlation function is of the SCS form. We
confirmed this result numerically for a particular CD model, the
CD4 model, which is defined below.\\

\begin{figure}[t!]
\includegraphics[width=6.0 cm,angle=-90,]{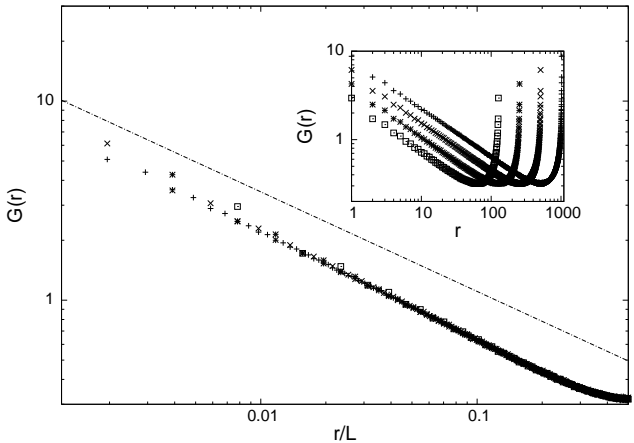}
\caption{Scaling of correlation functions of the VCD model. The inset
shows $G(r,L)$ versus $r$ for $L = 128, 256, 512$ and $1024$. The main figure
shows the scaling collapse when the same data are plotted versus $r/L$. The
straight line represents a power law with exponent $-0.5$. }
\label{scsfdpocorr}
\end{figure}

For the CD4 model, we consider a $1-d$ KPZ or EW surface, both of
whose surface configurations are known to be isomorphic to the
trajectories of a $1-d$ unbiased random walk, with the displacement of
the walk being the height of the surface.  For each configuration of
$2L$ lattice
sites, we fill the lattice up with $L$ particles, starting from the
bottommost site and moving upwards in height till all the particles
are exhausted (Figs. 1(a) and 1(b)). The filled sites are again
assigned a spin variable $+1$, and the unfilled sites $-1$. Thus, we
have again divided the lattice into two portions with the bottom half
filled with particles and the top half empty. While filling up, the
number of available sites at the topmost height generally exceeds the
number of particles that remain to be assigned.  To lift the
degeneracy, we assign particles randomly to the available sites. This
procedure is repeated over many configurations, and the results
averaged. Rather than dynamically evolving the surface, we drew
independent random walk trajectories so as to generate uncorrelated
surface configurations.\\

We monitored the two-point correlation function $C(r)$ of the CD4
model and found that it showed a behavior similar to Eq.~\ref{fdpocorr}
with $\alpha = 0.5$. Further, the probability distribution for the
length of the occupied and the unoccupied clusters was given by $P(l)\sim l^{-3/2}$,
as for the CD2 and CD3 models
discussed in the previous section. Thus, in common with these CD models, the CD4 model displays FDPO. The reason for choosing the CD4 model in the present study is that it leads to a VCD model with the desirable feature of a
strictly conserved number of sites and particles. Figure~\ref{scsfdpocorr} shows the result of the numerical simulation. We see that the two-point density-density correlation in the VCD model is a scaling function of separation $r$ and system size $L$ and that the scaling function diverges near the origin with an exponent $ \simeq 1/2$. This agrees well with the analytic prediction of Eq.~\ref{eqcdsps} on setting $\theta = 3/2$ and verifies the occurrence of SCS in the mapped VCD version of the model.\\

To summarize, we have shown a connection between fluctuation-dominated
phase ordereing and strong clustering states: Configurations of a CD
model, whose scaled correlations show a cusp singularity of the FDPO
type, can be transformed, via a simple mapping, into configurations of
a system with multiparticle occupancies, whose scaled correlations
show a divergence of a SCS variety.

\end{document}